\documentclass[a4paper,12pt]{article}
\usepackage{graphicx}
\begin{document}

\newcommand{\be}{\begin{equation}}
\newcommand{\beq}{\begin{equation}}
\newcommand{\eeq}{\end{equation}}
\newcommand{\ee}{\end{equation}}

\newcommand{\beqn}{\begin{eqnarray}}
\newcommand{\eeqn}{\end{eqnarray}}
\newcommand{\bea}{\begin{eqnarray}}
\newcommand{\ena}{\end{eqnarray}}
\newcommand{\ra}{\rightarrow}
\newcommand{\susy}{{{\cal SUSY}$\;$}}
\newcommand{\su}{$ SU(2) \times U(1)\,$}

\newcommand{\gag}{$\gamma \gamma$ }
\newcommand{\gagt}{\gamma \gamma }
\newcommand{\gam}{\gamma \gamma }
\def\W{{\mbox{\boldmath $W$}}}
\def\B{{\mbox{\boldmath $B$}}}
\def\V{{\mbox{\boldmath $V$}}}
\newcommand{\np}{Nucl.\,Phys.\,}
\newcommand{\pl}{Phys.\,Lett.\,}
\newcommand{\pr}{Phys.\,Rev.\,}
\newcommand{\prl}{Phys.\,Rev.\,Lett.\,}
\newcommand{\prep}{Phys.\,Rep.\,}
\newcommand{\zp}{Z.\,Phys.\,}
\newcommand{\sovjnp}{{\em Sov.\ J.\ Nucl.\ Phys.\ }}
\newcommand{\nuclinst}{{\em Nucl.\ Instrum.\ Meth.\ }}
\newcommand{\annp}{{\em Ann.\ Phys.\ }}
\newcommand{\intjmp}{{\em Int.\ J.\ of Mod.\  Phys.\ }}

\newcommand{\eps}{\epsilon}
\newcommand{\mw}{M_{W}}
\newcommand{\mww}{M_{W}^{2}}
\newcommand{\mwmw}{M_{W}^{2}}
\newcommand{\mhmh}{M_{H}^2}
\newcommand{\mz}{M_{Z}}
\newcommand{\mzz}{M_{Z}^{2}}

\newcommand{\cw}{\cos\theta_W}
\newcommand{\sw}{\sin\theta_W}
\newcommand{\tw}{\tan\theta_W}
\def\cww{\cos^2\theta_W}
\def\sww{\sin^2\theta_W}
\def\tww{\tan^2\theta_W}

\newcommand{\epm}{$e^{+} e^{-}\;$}
\newcommand{\epemt}{$e^{+} e^{-}\;$}
\newcommand{\epem}{e^{+} e^{-}\;}
\newcommand{\ememt}{$e^{-} e^{-}\;$}
\newcommand{\emem}{e^{-} e^{-}\;}

\newcommand{\lra}{\leftrightarrow}
\newcommand{\tr}{{\rm Tr}}
\def\ls1{{\not l}_1}
\newcommand{\cms}{centre-of-mass\hspace*{.1cm}}


\newcommand{\dkg}{\Delta \kappa_{\gamma}}
\newcommand{\dkz}{\Delta \kappa_{Z}}
\newcommand{\dz}{\delta_{Z}}
\newcommand{\dgz}{\Delta g^{1}_{Z}}
\newcommand{\dgzt}{$\Delta g^{1}_{Z}\;$}
\newcommand{\la}{\lambda}
\newcommand{\lag}{\lambda_{\gamma}}
\newcommand{\lambdae}{\lambda_{e}}
\newcommand{\laz}{\lambda_{Z}}
\newcommand{\lnl}{L_{9L}}
\newcommand{\lnr}{L_{9R}}
\newcommand{\lt}{L_{10}}
\newcommand{\lu}{L_{1}}
\newcommand{\ld}{L_{2}}
\newcommand{\eeww}{e^{+} e^{-} \ra W^+ W^- \;}
\newcommand{\eewwt}{$e^{+} e^{-} \ra W^+ W^- \;$}
\newcommand{\epemww}{e^{+} e^{-} \ra W^+ W^- }
\newcommand{\epemwwt}{$e^{+} e^{-} \ra W^+ W^- \;$}
\newcommand{\eennhht}{$e^{+} e^{-} \ra \nu_e \bar \nu_e HH\;$}
\newcommand{\eennhh}{e^{+} e^{-} \ra \nu_e \bar \nu_e HH\;}
\newcommand{\ppwg}{p p \ra W \gamma}
\newcommand{\wwhh}{W^+ W^- \ra HH\;}
\newcommand{\wwhht}{$W^+ W^- \ra HH\;$}
\newcommand{\ppwz}{pp \ra W Z}
\newcommand{\ppwgt}{$p p \ra W \gamma \;$}
\newcommand{\ppwzt}{$pp \ra W Z \;$}
\newcommand{\gamgamt}{$\gamma \gamma \;$}
\newcommand{\gamgam}{\gamma \gamma \;}
\newcommand{\egamt}{$e \gamma \;$}
\newcommand{\egam}{e \gamma \;}
\newcommand{\gamgamwwt}{$\gamma \gamma \ra W^+ W^- \;$}
\newcommand{\gamgamwwht}{$\gamma \gamma \ra W^+ W^- H \;$}
\newcommand{\gamgamwwh}{\gamma \gamma \ra W^+ W^- H \;}
\newcommand{\gamgamwwhht}{$\gamma \gamma \ra W^+ W^- H H\;$}
\newcommand{\gamgamwwhh}{\gamma \gamma \ra W^+ W^- H H\;}
\newcommand{\ggww}{\gamma \gamma \ra W^+ W^-}
\newcommand{\ggwwt}{$\gamma \gamma \ra W^+ W^- \;$}
\newcommand{\ggwwht}{$\gamma \gamma \ra W^+ W^- H \;$}
\newcommand{\ggwwh}{\gamma \gamma \ra W^+ W^- H \;}
\newcommand{\ggwwhht}{$\gamma \gamma \ra W^+ W^- H H\;$}
\newcommand{\ggwwhh}{\gamma \gamma \ra W^+ W^- H H\;}
\newcommand{\ggwwz}{\gamma \gamma \ra W^+ W^- Z\;}
\newcommand{\ggwwzt}{$\gamma \gamma \ra W^+ W^- Z\;$}

\newcommand{\ptu}{p_{1\bot}}
\newcommand{\vecptu}{\vec{p}_{1\bot}}
\newcommand{\ptd}{p_{2\bot}}
\newcommand{\vecptd}{\vec{p}_{2\bot}}
\newcommand{\ie}{{\em i.e.}}
\newcommand{\cm}{{{\cal M}}}
\newcommand{\cl}{{{\cal L}}}
\newcommand{\cd}{{{\cal D}}}
\newcommand{\cv}{{{\cal V}}}
\def\slashE{E\kern -.600em {/}}
\def\slashc{c\kern -.400em {/}}
\def\slashp{p\kern -.400em {/}}
\def\slashL{L\kern -.450em {/}}
\def\slashcl{\cl\kern -.600em {/}}
\def\Ww{{\mbox{\boldmath $W$}}}
\def\B{{\mbox{\boldmath $B$}}}
\def\noi{\noindent}
\def\nn{\noindent}
\def\sm{${\cal{S}} {\cal{M}}\;$}
\def\smn{${\cal{S}} {\cal{M}}$}
\def\nph{${\cal{N}} {\cal{P}}\;$}
\def\sb{$ {\cal{S}}  {\cal{B}}\;$}
\def\ssb{${\cal{S}} {\cal{S}}  {\cal{B}}\;$}
\def\ssbe{{\cal{S}} {\cal{S}}  {\cal{B}}}
\def\cviol{${\cal{C}}\;$}
\def\pviol{${\cal{P}}\;$}
\def\cpviol{${\cal{C}} {\cal{P}}\;$}

\newcommand{\lgg}{\lambda_1\lambda_2}
\newcommand{\lww}{\lambda_3\lambda_4}
\newcommand{\ppin}{ P^+_{12}}
\newcommand{\pmin}{ P^-_{12}}
\newcommand{\ppout}{ P^+_{34}}
\newcommand{\pmout}{ P^-_{34}}
\newcommand{\sinsq}{\sin^2\theta}
\newcommand{\cossq}{\cos^2\theta}
\newcommand{\yt}{y_\theta}
\newcommand{\hppll}{++;00}
\newcommand{\hpmll}{+-;00}
\newcommand{\hpplt}{++;\lambda_30}
\newcommand{\hpmlt}{+-;\lambda_30}
\newcommand{\hpptt}{++;\lambda_3\lambda_4}
\newcommand{\hpmtt}{+-;\lambda_3\lambda_4}
\newcommand{\dk}{\Delta\kappa}
\newcommand{\klam}{\Delta\kappa \lambda_\gamma }
\newcommand{\kac}{\Delta\kappa^2 }
\newcommand{\lac}{\lambda_\gamma^2 }
\def\gamgamtzz{$\gamma \gamma \ra ZZ \;$}
\def\gamgamtww{$\gamma \gamma \ra W^+ W^-\;$}
\def\gamgamtwwe{\gamma \gamma \ra W^+ W^-}

\def\sinb{\sin\beta}
\def\cosb{\cos\beta}
\def\sinbb{\sin (2\beta)}
\def\cosbb{\cos (2 \beta)}
\def\tgb{\tan \beta}
\def\tgbt{$\tan \beta\;\;$}
\def\tgbsq{\tan^2 \beta}
\def\sinal{\sin\alpha}
\def\cosal{\cos\alpha}
\def\stop{\tilde{t}}
\def\sto{\tilde{t}_1}
\def\stt{\tilde{t}_2}
\def\stl{\tilde{t}_L}
\def\str{\tilde{t}_R}
\def\msto{m_{\sto}}
\def\mstosq{m_{\sto}^2}
\def\mstt{m_{\stt}}
\def\msttsq{m_{\stt}^2}
\def\mt{m_t}
\def\mtsq{m_t^2}
\def\sint{\sin\theta_{\stop}}
\def\sintt{\sin 2\theta_{\stop}}
\def\cost{\cos\theta_{\stop}}
\def\sintsq{\sin^2\theta_{\stop}}
\def\costsq{\cos^2\theta_{\stop}}
\def\mqtt{\M_{\tilde{Q}_3}^2}
\def\mutt{\M_{\tilde{U}_{3R}}^2}
\def\sbottom{\tilde{b}}
\def\sbo{\tilde{b}_1}
\def\sbt{\tilde{b}_2}
\def\sbl{\tilde{b}_L}
\def\sbr{\tilde{b}_R}
\def\msbo{m_{\sbo}}
\def\msbosq{m_{\sbo}^2}
\def\msbt{m_{\sbt}}
\def\msbtsq{m_{\sbt}^2}
\def\mt{m_t}
\def\mtsq{m_t^2}
\def\selectron{\tilde{e}}
\def\seo{\tilde{e}_1}
\def\set{\tilde{e}_2}
\def\sel{\tilde{e}_L}
\def\ser{\tilde{e}_R}
\def\mseo{m_{\seo}}
\def\mseosq{m_{\seo}^2}
\def\mset{m_{\set}}
\def\msetsq{m_{\set}^2}
\def\msel{m_{\sel}}
\def\mser{m_{\ser}}
\def\me{m_e}
\def\mesq{m_e^2}
\def\snu{\tilde{\nu}}
\def\snue{\tilde{\nu_e}}
\def\set{\tilde{e}_2}
\def\snul{\tilde{\nu}_L}
\def\msnue{m_{\snue}}
\def\msnuesq{m_{\snue}^2}
\def\smuon{\tilde{\mu}}
\def\smul{\tilde{\mu}_L}
\def\smur{\tilde{\mu}_R}
\def\msmul{m_{\smul}}
\def\msmulsq{m_{\smul}^2}
\def\msmur{m_{\smur}}
\def\msmursq{m_{\smur}^2}
\def\stau{\tilde{\tau}}
\def\stauo{\tilde{\tau}_1}
\def\staut{\tilde{\tau}_2}
\def\staul{\tilde{\tau}_L}
\def\staur{\tilde{\tau}_R}
\def\mstauo{m_{\stauo}}
\def\mstauosq{m_{\stauo}^2}
\def\mstaut{m_{\staut}}
\def\mstautsq{m_{\staut}^2}
\def\mtau{m_\tau}
\def\mtausq{m_\tau^2}
\def\gluino{\tilde{g}}
\def\mgluino{m_{\tilde{g}}}
\def\mchi{m_\chi^+}
\def\neuto{\tilde{\chi}_1^0}
\def\mneuto{m_{\tilde{\chi}_1^0}}
\def\neutt{\tilde{\chi}_2^0}
\def\mneutt{m_{\tilde{\chi}_2^0}}
\def\neutth{\tilde{\chi}_3^0}
\def\mneutth{m_{\tilde{\chi}_3^0}}
\def\neutf{\tilde{\chi}_4^0}
\def\mneutf{m_{\tilde{\chi}_4^0}}
\def\chargop{\tilde{\chi}_1^+}
\def\mchargo{m_{\tilde{\chi}_1^+}}
\def\chargtp{\tilde{\chi}_2^+}
\def\mchargt{m_{\tilde{\chi}_2^+}}
\def\chargom{\tilde{\chi}_1^-}
\def\chargtm{\tilde{\chi}_2^-}
\def\bino{\tilde{b}}
\def\wino{\tilde{w}}
\def\photino{\tilde{\gamma}}
\def\zino{tilde{z}}
\def\sdowno{\tilde{d}_1}
\def\sdownt{\tilde{d}_2}
\def\sdownl{\tilde{d}_L}
\def\sdownr{\tilde{d}_R}
\def\supo{\tilde{u}_1}
\def\supt{\tilde{u}_2}
\def\supl{\tilde{u}_L}
\def\supr{\tilde{u}_R}
\def\mh{m_h}
\def\mht{m_h^2}
\def\MH{M_H}
\def\MHt{M_H^2}
\def\MA{M_A}
\def\MAt{M_A^2}
\def\MHp{M_H^+}
\def\MHm{M_H^-}

\begin{titlepage}
\def\baselinestretch{1.2}
\begin{center}
{\large {\bf {\em The MSSM invisible Higgs in the light of  dark
matter and $g-2$ .}}}

\vspace*{1.cm}

\begin{tabular}[t]{c}

{\bf G.~B\'elanger$^{1}$, F.~Boudjema$^{1}$, A.~Cottrant$^{1}$,
R.~M.~Godbole$^{2}$ and A.~Semenov$^{1}$}
\\
\\
\\
{\it 1. Laboratoire de Physique Th\'eorique} {\large
LAPTH}$^\dagger$\\
 {\it Chemin de Bellevue, B.P. 110, F-74941 Annecy-le-Vieux,
Cedex, France.}\\

 {\it 2. Centre for Theoretical Studies, Indian
Institute of Science}
\\ {\it Bangalore 560 012, India }

\end{tabular}
\end{center}

\centerline{ {\bf Abstract} } \baselineskip=14pt \noindent
{\small Giving up the assumption of the gaugino mass unification
at the GUT scale, the latest LEP and Tevatron data  still allow
the lightest supersymmetric Higgs to have a large branching
fraction into invisible neutralinos. Such a Higgs may be difficult
to discover at the LHC and is practically unreachable at the
Tevatron. We argue that, for some of these  models to be
compatible with the relic density,  light sleptons with masses not
far above the current limits are needed. There are however models
that allow for larger sleptons masses without being in conflict
with the relic density constraint. This is possible because these
neutralinos can annihilate efficiently through a Z-pole. We also
find that many of  these models can nicely account, at the
$2\sigma$ level, for the discrepancy in the latest $g-2$
measurement. However, requiring  consistency with the $g-2$ at the
$1 \sigma$ level, excludes models that lead to the largest Higgs
branching fraction into LSP's. In all cases one expects that even
though the Higgs might escape detection, one would have a rich
SUSY phenomenology even at the Tevatron, through the production of
 charginos and neutralinos.}
\vspace*{\fill}

\vspace*{0.1cm} \rightline{LAPTH-856/2001}
\vspace*{0.1cm}\rightline{IISc-CTS/14/01}

\rightline{{\large June 2001}} \vspace*{1cm} $^\dagger$URA 14-36
du CNRS, associ\'ee  \`a l'Universit\'e de Savoie.
\end{titlepage}
\baselineskip=18pt


\section{Introduction}
With the naturalness argument, the latest electroweak data that
suggest a light Higgs make supersymmetry the most probable
candidate for New Physics especially as it can also solve the dark
matter problem. In most scenarios the lightest supersymmetric
particle is a neutral, stable, weakly interacting particle: the
neutralino LSP. Current limits\cite{leplimit2001} on both the
Higgs and the neutralino in a general SUSY model are such that it
is kinematically possible for the Higgs to decay into the lightest
neutralino. If the decay rate is substantial the Higgs will be
mainly invisible, while its usual branching ratios will be
dramatically reduced preventing a detection in the much studied
channels at the LHC and the Tevatron. Some theoretical
studies\cite{Kane-invisible-lhc,DP-invisible-lhc,Gunion-invisible-lhc,Tevatron-invisible-lhc,Zeppenfeld-h-invisible}
have addressed the issue of how to hunt an invisibly decaying
Higgs at a hadronic machine, with optimistic conclusions
especially in the case of the LHC. At the
Tevatron\cite{Tevatron-invisible-lhc} requiring a $5 \sigma$
discovery of an invisible Higgs with as much as $100\%$ branching
into invisibles, ${\rm BR}_{\rm inv}$, will need more than
$30fb^{-1}$ for a Higgs mass consistent with the direct limit from
LEP. Therefore the prospect for the detection of an invisibly
decaying Higgs at the Tevatron seems dim. As for the LHC it has
been suggested to use $WH/ZH$ production which could be efficient
if ${\rm BR}_{\rm inv}>25\%$ with a luminosity of
$100fb^{-1}$\cite{DP-invisible-lhc}, while $t\bar t
h$\cite{Gunion-invisible-lhc} would require ${\rm BR}_{\rm
inv}>60\%$. Both these studies should be updated and are in need
of a full simulation. A recent
suggestion\cite{Zeppenfeld-h-invisible} has been to exploit the
$W$ fusion process. The results for the latter are quite promising
since for a luminosity of $100fb^{-1}$ a branching ratio into
invisibles as low as $5\%$ is enough for Higgs discovery. It rests
that  a full simulation that should  tackle the issue of trigger
is needed, before one draws definite conclusions. The aim of the
present study is to find out how large the branching ratio into
neutralinos can be, taking into account the present data and also
what accompanying SUSY phenomenology, if any, should we be
prepared to look for in such eventuality. This letter is an update
and an extension of a comprehensive study we have made
recently\cite{nous_hinvisible_lhc}. Since we will be dealing with
a rather light SUSY spectrum we will here also include a
discussion about the latest limit on the muon $g-2$ from the E821
experiment\cite{g-2ex} and whether the scenarios we are
considering help account for the reported discrepancy with the \sm
value.

Our starting point is to find out under which conditions a large
invisible width of the Higgs due to neutralinos is possible. The
width of the lightest Higgs to the lightest neutralinos
writes\cite{Haber-old-invisible}
\beqn
\label{gamhtolsp} \Gamma(h \ra \neuto \neuto)&=&\frac{G_F \mw
\mh}{2 \sqrt{2} \pi} \;(1-4 \mneuto^2/m_h^2)^{3/2}\;  |C_{h \neuto
\neuto}|^2 \nonumber \\
{\rm where \;\;\;\;} C_{h \neuto \neuto}&=&(O^N_{12}-\tw
O^N_{11})(\sinal\; O^N_{13}\;+\; \cosal \; O^N_{14}) \nonumber
\\  &\simeq&
(O^N_{12}-\tw O^N_{11})(\sinb\; O^N_{14}\;- \; \cosb
\;O^N_{13})\;\;{\rm for} \;\; M_A \gg M_Z
\eeqn

$O^N_{ij}$ are the elements of the orthogonal (we assume \cpviol
conservation) matrix which diagonalizes the neutralino mass matrix
(for convention and definition, see \cite{nous_hinvisible_lhc}) .
$\alpha$ is the angle that enters the diagonalization of the
\cpviol-even neutral Higgses which in the decoupling limit (large
$M_A$ and ignoring radiative corrections) is trivially related to
the angle $\beta$. $|O^{N}_{1j}|^2$ defines the composition of the
lightest neutralino $\neuto$. $j=1$ defines the bino component,
$j=2$ the wino, while $j=3,4$ give the Higgsino component. It is
clear then, apart from phase space, that the LSP has to be a
mixture of gaugino and higgsino in order to have a large enough
coupling to the Higgs. Since the lightest MSSM Higgs mass can not
exceed $135$GeV, one must require the LSP to be lighter than about
$65$GeV. This puts rather strict constraints on $M_2$ and $\mu$,
since these parameters also define the chargino masses whose limit
is about $103$GeV\cite{leplimit2001}, almost independently of any
other SUSY parameter. Thus one needs $M_1$ to be small enough so
that it sets the mass of the neutralino which will then be, to a
large degree, a bino. However one can not make $\mu$ too small
either, otherwise one washes out any higgsino component which is
essential to get enough mixing for the neutralino to couple to the
Higgs.  The fact that one tries to make $\mu$ as small as possible
means that large mixings entail also light charginos and
neutralinos NLSP not far above the present experimental limit. One
also finds \cite{nous_hinvisible_lhc} that positive $\mu$ values
are preferred. One would think that by taking larger values of
$\tgb$ one would make the Higgs mass higher which will allow more
phase space for the invisible decay. However, we find
\cite{nous_hinvisible_lhc} that the LSP masses increase  even
faster  and their coupling to the Higgs gets smaller with
increasing $\tgb$. Therefore the largest effects for the invisible
Higgs occur for moderate $\tgb$.

Most collider constraints on the neutralino refer to the so-called
gaugino unification condition $M_1=\frac{5}{3} \tan^2\theta_W M_2
\simeq M_2/2$. In this case the limit on the lightest neutralino
is set by the chargino which in turn leaves very little room for
an appreciable Higgs decay into invisible neutralinos. In our
previous paper\cite{nous_hinvisible_lhc} we found that, for such
models, this branching is never above $20\%$ and thus does not
endanger the searches in the conventional channels. Previously we
had concentrated on the case $M_1=M_2/10$ valid at the electroweak
scale and allowed $M_2$ and $\mu$ to vary.  Though the value of
$M_1$ with respect to $\mu$ was not optimised, substantial
branching into invisible was found. In the present analysis we
seek a larger higgsino-gaugino mixing and instead of $M_1=M_2/10$
we also study $M_1=M_2/5$, at the electroweak scale, in detail. We
thus also allow for larger LSP masses which, as we will see, lead
to some quite interesting novel features especially as concerns
cosmological considerations. We will also investigate which range
of $M_1$, independently of $M_2$ and $\mu$, give the largest
invisible branching ratio.

\section{MSSM models for an invisible Higgs and constraints}
Our scenario requires as large a Higgs mass as possible without
making \tgbt too large. We will then only consider the MSSM in the
decoupling limit with $M_A \sim 1$TeV and choose large enough stop
masses ($m_{\tilde{t}}=$1TeV) and large mixing ($A_t=2.4$TeV).
With $\tgb>5$, we could essentially consider the Higgs mass as a
free parameter. We have imposed $m_h>113$GeV and with our
parameters we have $m_h=125$GeV ($128$GeV) for $\tgb=5$(10).

The limits on $M_1,M_2,\mu$, the key ingredients for this
analysis, are set from the chargino mass limit at LEP2,
$m_{\chi_1^\pm}>103$GeV\cite{leplimit2001}. This bound can be
slightly relaxed depending on $\tgb$ and the sneutrino mass,
however we prefer to take the strongest constraint so that our
results are more robust. The cross section into neutralinos at
LEP2, $\sigma (\epem \ra \neuto \neutt + \neuto \neutth)$ could in
principle also help reduce the parameter space for these
non-unified gaugino mass models. The neutralino cross section
constraint,  as opposed to the chargino mass limit, depends
crucially on the higgsino content of the produced neutralinos, as
well as on the mass of the selectron and the decay pattern. We
impose $\sigma (\epem \ra \neuto \neutt +\neuto \neutth \ra
\slashE \mu^+ \mu^-)<.1pb$, for $\sqrt{s}=208$GeV. Our formulae
for the branching ratios of the heavier neutralinos include all
two and three body decays. For the parameters we have studied we
find, in fact, that this constraint does not overcome the chargino
mass limit. We have also imposed the limits on the invisible width
of the $Z$\cite{LEPelectroweak}:
\beqn
\label{invisibleZ} \Gamma^Z_{{\rm inv}}\equiv \Gamma(Z\ra \neuto
\neuto) < 3 MeV
\eeqn

We will also take $m_{\tilde{l}}>96$GeV, for all sleptons $\tilde
l$, even though the limit on the lightest stau is slightly
lower\cite{leplimit2001}.

Scenarios with low $M_1$ that have very light neutralino LSP into
which the Higgs can decay, suppressing quite strongly its visible
modes,  can contribute quite substantially to the relic density
$\Omega h^2$,  if all sfermions are heavy. Indeed, in the models
we are considering the LSP is {\em mainly} (but not totally) a
bino. Since it is rather light  the annihilation channels are into
the light fermions and therefore the largest contributions are
from processes involving ``right-handed" sleptons. This is because
the latter have the largest hypercharge. In this case the relic
density may be approximated as $\Omega h^2 \sim 10^{-3}
m_{\tilde{l}_R}^4/ \mneuto^2$ (all masses in GeV) which shows how
the  strong constraint on $m_{\tilde{l}_R}$ rapidly sets in.
However this limit can become irrelevant in the models we
consider. Interestingly, allowing larger neutralino masses than in
our previous analysis\cite{nous_hinvisible_lhc}, annihilation
through the $Z$ pole, $\neuto \neuto \ra Z$, can become very
effective. The above formula for the relic density no longer holds
then. We use a new code\cite{OmegaComphep} for the calculation of
the relic density that tackles all $s$-channels poles, threshold
effects and includes all co-annihilations channels (including
slepton, neutralino and chargino co-annihilations). The program
extracts all {\em exact} matrix elements (for about 500 processes)
from {\tt CompHEP}\cite{comphep} and is linked to {\tt
HDECAY}\cite{HDECAY} and {\tt FeynHiggs}\cite{FeynHiggs} for the
Higgs sector. When possible, checks against {\tt
DarkSUSY}\cite{DarkSUSY} have been performed. The agreement is
generally quite good. In the last few years constraints on the
cosmological parameters that enter the calculation of the relic
density have improved substantially. Various
observations\cite{omega} suggest to take as a benchmark $ \Omega
h^2 <.3$ where we identify $\Omega$ with  the fraction of the
critical energy density provided by neutralinos. $h$ is the Hubble
constant in units of 100 km sec$^{-1}$ Mpc$^{-1}$. This constraint
is  consistent with limits on the age of the
Universe\cite{hubble}, the measurements of the lower multipole
moment power spectrum from CMB data and the determination of
$\Omega_{\rm matter}$ from rich clusters, see \cite{omega} for
reviews. It also, {\em independently}, supports data from type Ia
supernovae\cite{supernova} indicative for a cosmological constant.
Note that it is not essential to impose the lower bound $\Omega
h^2 >.1$. A lower value of $\Omega h^2$ would mean that one needs
other form of dark matter than the SUSY models one is considering.
Our bound $\Omega h^2 <.3$ can be considered as quite conservative
in view of the latest CMB data from BOOMERANG\cite{Boomerang2001},
MAXIMA\cite{Maxima2001} and Dasi\cite{Dasi2001}. The latter
extracts $\Omega h^2=.14\;\pm\;.04$ almost independently of the
choice of a ``prior" on $h$ and thus the $2\sigma$ upper bound is
$.22$. This is also consistent with the latest BOOMERANG data with
a very weak ``prior" on $h$, $.45<h<.9$ and the requirement of an
Universe older than $10$Gyr\cite{hubble}. Combining this with
stronger priors including type Ia supernovae\cite{supernova} and
analysis of Large Scale Structure (LSS)\cite{LSSanalysis} together
with the theoretical bias $\Omega_{\rm tot}=1$, gives the rather
precise constraint $\Omega h^2=.13\;\pm\;.01$, which at $2\sigma$
would only allow $\Omega h^2<.15$ for any SUSY contribution.
Although one should be cautious at this stage about using such a
strict  bound considering that some of the cosmological parameters
are still subject to fluctuations, we will comment briefly on how
our results change if one takes this strict constraint at face
value.

For the calculation of the relic density one needs a model for the
SUSY masses. We assumed all squarks to be heavy. In any case
squarks compared to ``right sleptons" do not contribute much to
the annihilation cross section for a bino LSP. On the other hand
heavy squarks, especially stops would be required in order to get
a heavy enough light Higgs. A simple model would be to take a
common scalar mass $m_0$ (defined at the GUT scale) for the SUSY
breaking sfermion mass terms of both left and right sleptons of
all three generations. As for the gaugino masses, to obtain $M_1=r
M_2$ at the electroweak scale one needs $\bar{M}_1 \simeq 2 r
\bar{M}_2 $ at the GUT scale. $\bar{M}_2$ is the $SU(2)$ gaugino
mass at the GUT scale which again relates to $M_2$ at the
electroweak scale as $M_2 \sim 0.825 \bar {M}_2$. For $r<1/3$ or
so, this scheme leads to almost no running of the right slepton
mass, since the contribution from the running is of order $M_1^2$,
while left sleptons have an added $M_2^2$ contribution and would
then be ``much heavier". Indeed, neglecting Yukawa couplings one
may write, with $M_1=r M_2$ at the electroweak scale

\beqn
\label{m0running} \mser^2&=&m_0^2\;+\; .88 \;r^2 M_2^2\;-\;\sww
D_z \nonumber
\\ \msel^2&=&m_0^2\;+ (0.72+.22\; r^2) M_2^2\; -\;(.5-\sww)D_z \nonumber
\\ \msnue^2&=&m_0^2\;+ (0.72+.22 \;r^2) M_2^2\;+\;D_z/2
\;\;\;\;\;\; {\rm with} \nonumber
\\  D_z&=&\mzz \cosbb
\eeqn

\noi Note that squarks can be made much heavier than the sleptons
even by taking the same common scalar mass  since they receive a
large contribution from the SU(3) gaugino mass. Of course, to
allow for a low $\mu$ in this scenario one needs to appropriately
choose the soft SUSY Higgs scalar masses at high scale. It is
important to stress that the kind of models we investigate in this
letter are quite plausible. The GUT-scale relation which equates
all the gaugino masses at high scale need not be valid in a more
general scheme of SUSY breaking. In fact even within SUGRA this
relation need not necessarily hold since it requires the kinetic
terms for the gauge superfields to be the most simple and minimal
possible (diagonal and equal). One can easily arrange for a
departure from equality by allowing for more general forms for the
kinetic terms\cite{nmSUGRA}.  Within $SU(5)$ this occurs when the
auxiliary component of a superfield transforms as a {\bf 24}
dimensional representation. In this case one gets $M_1=M_2/6$, at
the electroweak scale, but $M_3=2 M_2$\cite{nonuni-24}. In
superstring models, although dilaton dominated manifestations lead
to universal gaugino masses, moduli-dominated or a mixture of
moduli and dilaton fields lead also to non universality of the
gaugino masses\cite{nonuiniversal-strings} and may or may not
(multi-modulii\cite{multimoduli}) lead to universal scalar masses.
The so-called anomaly-mediated SUSY breaking
mechanisms\cite{non-universal-anomaly} are also characterised by
non-universal gaugino masses, though most models in the literature
lead rather to $r>1$ which is irrelevant for the Higgs search.

Since the model requires light charginos and since dark matter
argument may force us to also consider light sleptons one should
inquire whether these scenarios may account for the latest $g-2$
results\cite{g-2ex} from the E821 experiment at Brookhaven. First,
as stressed in our previous analyses\cite{nous_hinvisible_lhc}
models that lead to the largest branching into invisibles have
$\mu >0$, which is preferred by $g-2$. Though large \tgbt values
do give a larger $g-2$ they  do not give as large branching into
invisibles. Moderate \tgbt that give a large invisible Higgs decay
should also have light sleptons to account for $g-2$. We will
discuss the situation by imposing the $2\sigma$ limit, $1.1 \;
10^{-9} < a_{\mu}^{\rm susy}< 7.5 \; 10^{-9}$, on $g-2$ as well as
what remains when one does not take into account the observed
discrepancy in the measurement of $g-2$. Our calculation of $g-2$,
which we have checked against some of the computations in the
extensive literature\cite{g-2susy}, includes also the effect of
$A_\mu$, the tri-linear soft-Susy breaking parameter in the smuon
sector. However all of our discussion refers to the situation with
$A_\mu=0$. We have checked that especially in the regions that
lead to the largest branching into invisibles, the results are not
much dependent on $A_\mu$.

\begin{figure*}[htbp]
\begin{center}
\includegraphics[width=16cm,height=18cm]{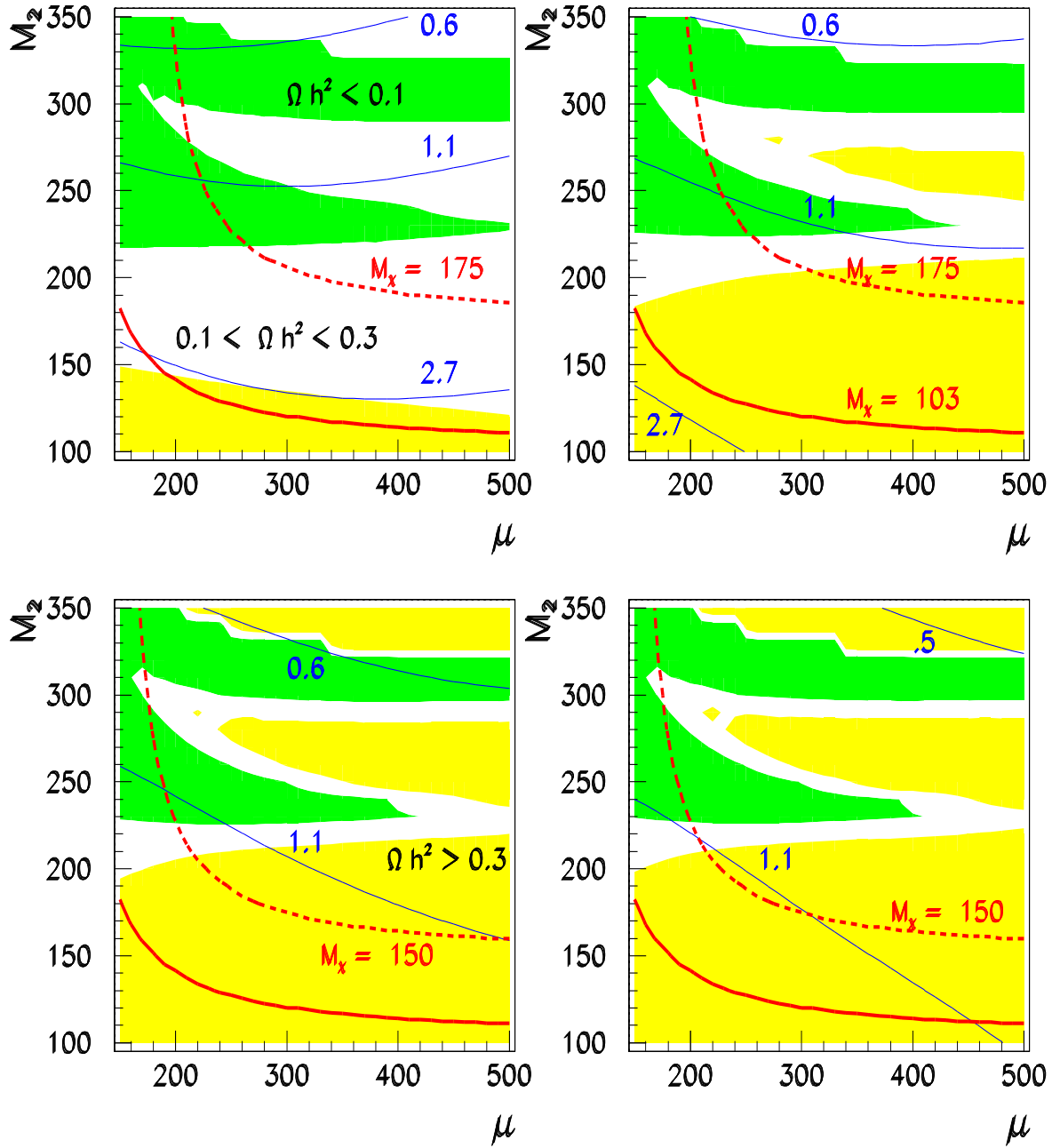}
\caption{\label{fig-constraints}{\em Constraints on the parameter
space for $\tgb=5$ and $M_1=M_2/5$, for four values of the slepton
mass $m_0=100,140,180,220$GeV from left to right and top to
bottom. Slepton masses are defined via $m_0$ according to
Eq.~\ref{m0running}. The thick (red) line defines the chargino
mass constraint $m_{\chi^+}>103$GeV (the area below the line is
excluded). The dashed (red) line corresponds to
$m_{\chi^+}>175$GeV for $m_0=100,140$GeV and $m_{\chi^+}>150$GeV
for $m_0=180,220$GeV which we estimate (conservatively) as being
the Tevatron RunII reach. The light grey (yellow) area has $\Omega
h^2>.3$ and is therefore excluded. The dark grey area (green) has
$\Omega h^2<.1$. The white area is the cosmologically preferred
scenario with $.1<\Omega h^2<.3$. The thin (blue) lines are
constant $a_\mu$ lines in units of $10^{-9}$ so that $1.1$ ($2.7$)
corresponds to the $2\sigma$ ($1\sigma$) present lower bound\/.}}
\end{center}
\end{figure*}

Fig.~\ref{fig-constraints} shows the allowed parameter space in
the $M_2, \mu$ plane with $\tgb=5$ and $M_1=M_2/5$ for four
different values of $m_0$. To a good approximation $m_0$ can be
identified with the ``right slepton" mass. The chargino mass limit
from LEP2 is delimited by a line. It does not depend on $m_0$. The
direct LEP2 limits, expectedly, cut on the lowest $\mu,M_2$
region. This is in contrast to the relic density requirement which
depend sensitively on $m_0$. We delineate three regions set by the
relic density:a) the overclosure region $ \Omega h^2>.3$ which we
consider as being definitely ruled out, b) $ .1 <\Omega h^2<.3$
which is the preferred region and c) $ \Omega h^2<.1$ where there
is simply not enough susy dark matter. As $m_0$ increases the
allowed region for the relic density shrinks. These remaining
allowed regions correspond essentially to the pole annihilation
$\neuto \neuto \ra Z$. Also shown is the line corresponding to the
lower $2\sigma$ limit on the $g-2$, which becomes also  more
constraining as $m_0$ increases. To compensate for the increase in
$m_0$, smaller combinations of $\mu-M_2$ corresponding to lighter
charginos in the loop are picked up. It is worth stressing that
the $g-2$ measurement constrains regions with large $\mu-M_2$
values especially for large $m_0$, but these regions as, we will
see, do not correspond to the largest branching ratio into
invisibles. Note that, especially for this somewhat low value of
$\tgb$, we never find large contributions to $g-2$. In fact if one
slightly relaxes the $g-2$ limit by requiring $a_\mu>0$, one has
for the parameters of interest no constraint from $g-2$. On the
other hand, had we imposed, for $\tgb=5$, that the SUSY
contribution be within $1 \sigma$ we would not have found a
solution, apart from a tiny ``hole" at low $m_0=100$GeV. Finally,
we note that $b\ra s\gamma$ is irrelevant since the squarks and
gluinos are assumed heavy and that we are choosing $\mu>0$ anyway.

\section{Results}
The branching ratio into invisible due to neutralinos will be
denoted by $B_{\chi \chi}$. The opening up of this channel will
not have any effect on any of the Higgs production mechanisms.
This is in contrast to other SUSY effects on the production and
decay of the Higgs, like those due to a light stop, see for
instance \cite{nous_Rggstophiggs_lhc}. Thus the Higgs discovery
significances of the different channels at the LHC (and the
Tevatron) are only affected by the reduction in the branching
ratio into $b\bar b$ and $\gamma \gamma$. We define $R_{bb}$ as
the reduction factor of the branching ratio of $h \ra b \bar b$
due to invisible compared to the same branching ratio of a
standard model Higgs with the same Higgs mass:
\beqn \label{Rbb}
 R_{b b}=\frac{ BR^{SUSY}(h
\ra b \bar b)}{BR^{SM}(h \ra b \bar b)}
\eeqn
Likewise we define $R_{\gamma \gamma}$ for the branching ratio
into $\gamma \gamma$. Since in the absence of light neutralinos
the width of the Higgs is dominated by that into $b\bar b$, one
has roughly

\beqn
R_{b b} \sim R_{\gamma \gamma} \sim 1-B_{\chi \chi}
\eeqn

This is well supported by our full analysis and therefore we will
refrain from showing simultaneously the behaviour of all these
three observables .

\begin{figure*}[htbp]
\begin{center}
\includegraphics[width=16cm,height=18cm]{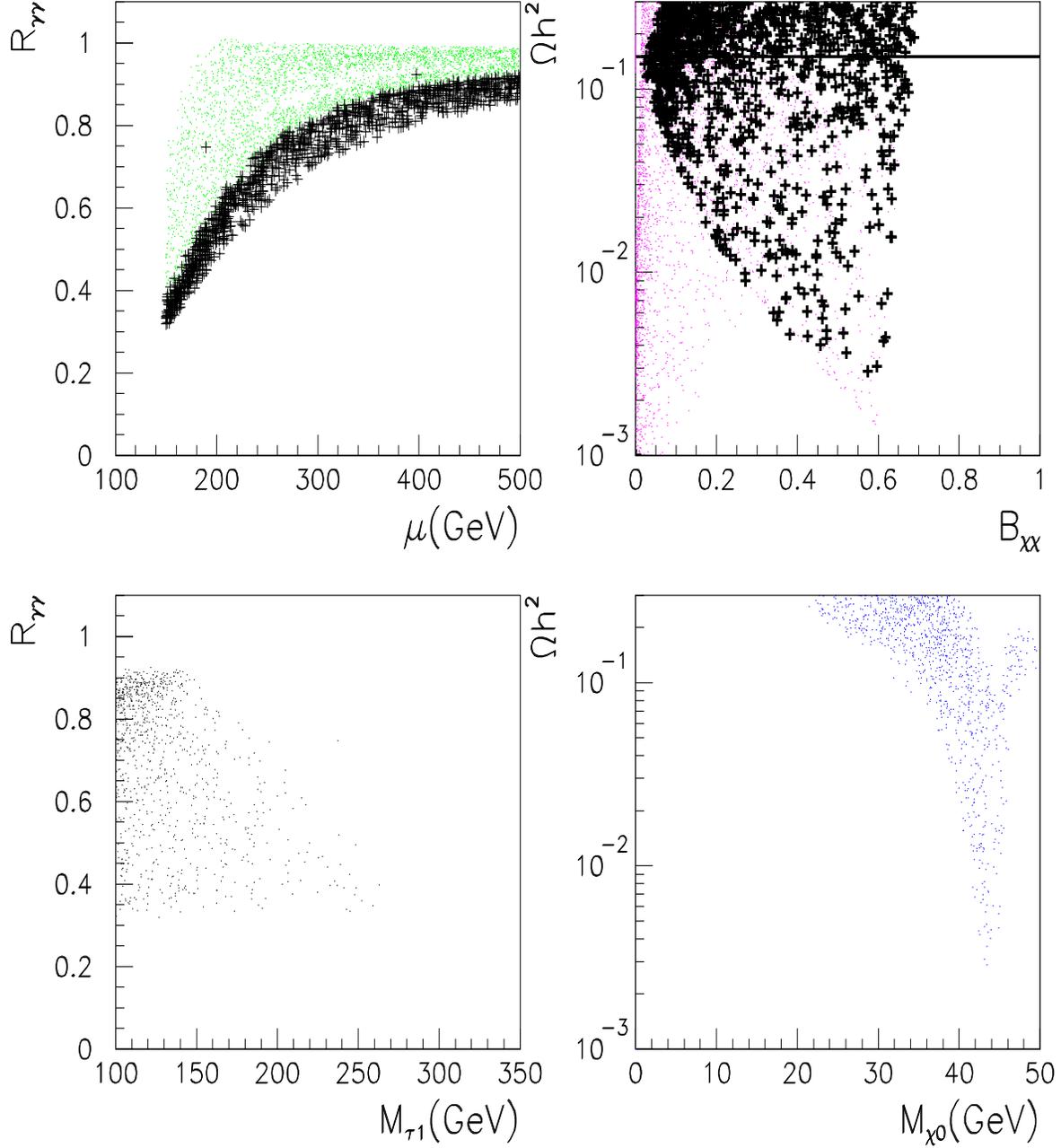}
\caption{\label{tgb5}{\em Results for $\tgb=5$ and $M_1=M_2/5$,
scanning over $M_2,\mu$ and $m_0$.  The first panel shows
$R_{\gamma \gamma}$ {\it vs.} $\mu$. The area with the crosses has
$g-2$ imposed at $2\sigma$ while the additional light shaded
(green) region does not have this constraint. The second panel
gives the branching ratio into invisibles {\it vs} the relic
density with $\Omega h^2<.3$. In the region with crosses the
$2\sigma$ $g-2$ constraint has been imposed while in the
additional area (pink) this constraint was removed. Also shown in
this panel by the (horizontal) line is the strict bound from
BOOMERANG with priors $\Omega h^2<.15$. The third panel (bottom
left) shows the correlation between the lightest slepton mass
($\tilde{\tau_1}$) and the drop in the two photon rate. The last
panel exhibits the annihilation through the $Z$ pole by showing
the behaviour of the relic density {\it vs} the mass of the
neutralino LSP. \/}}
\end{center}
\end{figure*}

\begin{figure*}[htbp]
\begin{center}
\includegraphics[width=16cm,height=16cm]{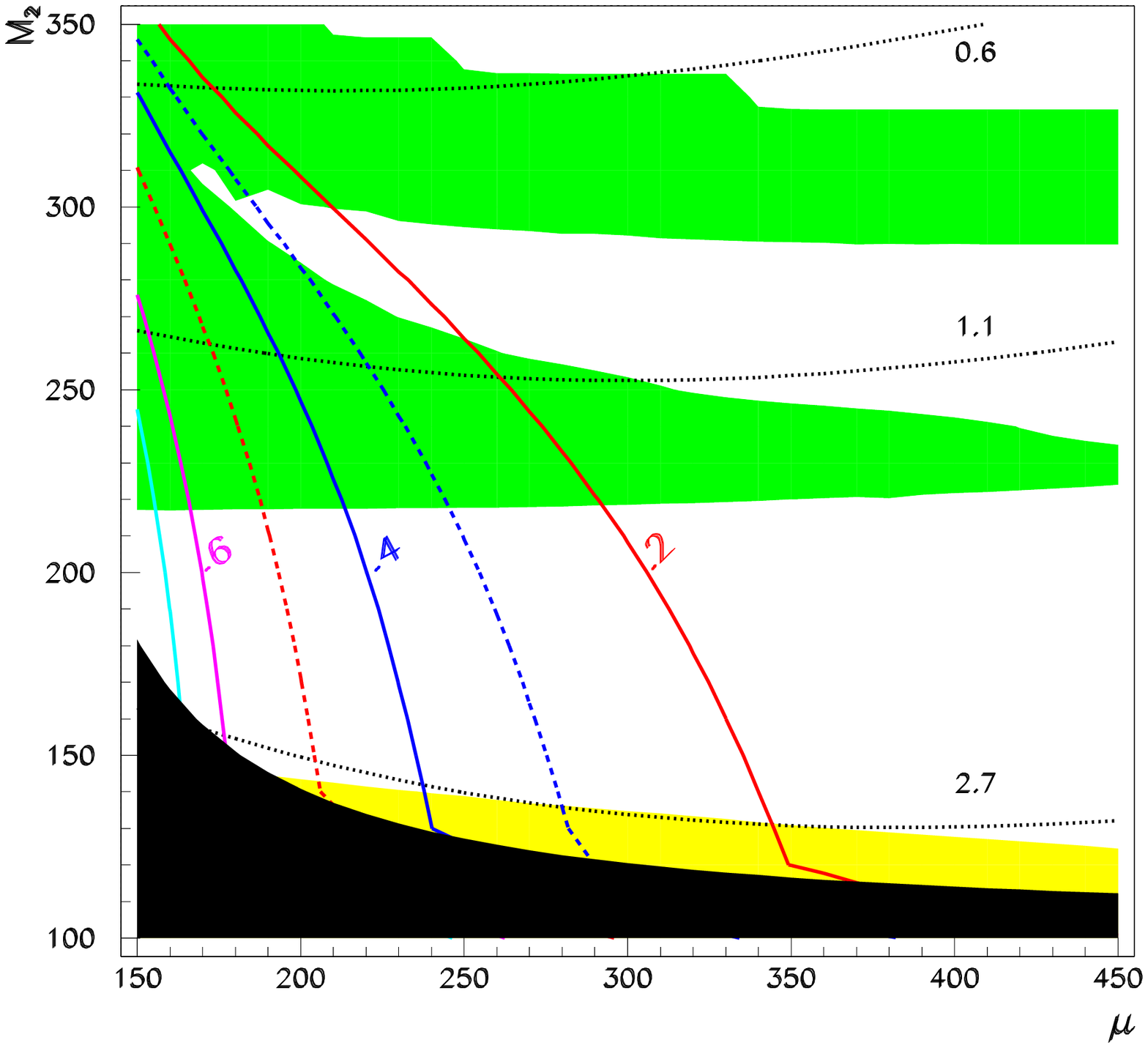}
\caption{\label{tgb5cont}{\em With the parameters as in the
previous figure, contours of constant $Br_{\chi \chi}$ from .2
(far right) to .65 (far left). We have also superimposed  the
various constraints, choosing $m_0=100$GeV, which correspond to
the first panel of Fig.~1. The black area is excluded by the
chargino mass at LEP. The other shadings refer to the relic
density (as in Fig.~1). The dotted lines are constant $a_\mu$
lines in units of $10^{-9}$\/.}}
\end{center}
\end{figure*}
We take a scenario with $\tgb=5$ and $M_1=M_2/5$ and scan over
$\mu,M_2, m_0$ (defined in Eq.~\ref{m0running}) with
$70<m_0<300$GeV, $100<M_2<350$GeV and $150<\mu<500$GeV. We see, in
Fig.~\ref{tgb5}, that indeed the largest drop in $R_{\gamma
\gamma}$ is for the lowest allowed value of $\mu$, which as argued
earlier maximises the higgsino component. The second panel of the
figure shows that even after putting the $g-2$ constraint, a large
fraction of the parameter space is compatible with the relic
density constraint, many models giving even just the needed amount
of dark matter, $.1 -.3$. One also sees that large values of
slepton masses are still compatible with dark matter and lead to
large drops in the channels with visible signatures. As the figure
clearly shows this is due to the efficient annihilation at the $Z$
pole. We also show (second panel), that imposing the strict bound
suggested by BOOMERANG, $\Omega h^2<.15$, still allows values of
$B_{\chi \chi}$ as large as about $70\%$. Fig.~3 shows the
different contours in the $M_2-\mu$ plane of $B_{\chi \chi}$
together with the constraint from the relic density and $g-2$. We
see that, even after  taking all these constraints, we still find
large branching ratio of the lightest SUSY Higgs into neutralinos
and we confirm that the largest branchings correspond to the
smallest $\mu$ values which are not terribly constrained by dark
matter and $g-2$. Insisting on explaining the $g-2$ value at
$1\sigma$ for $m_0=100$GeV selects  a tiny region corresponding to
$B_{\chi \chi}$ in the range $.4-.6$. It is also worth stressing
that even in these general models, the branching ratio into
invisible is never larger than $70\%$.

\begin{figure*}[htbp]
\begin{center}
\includegraphics[width=16cm,height=10cm]{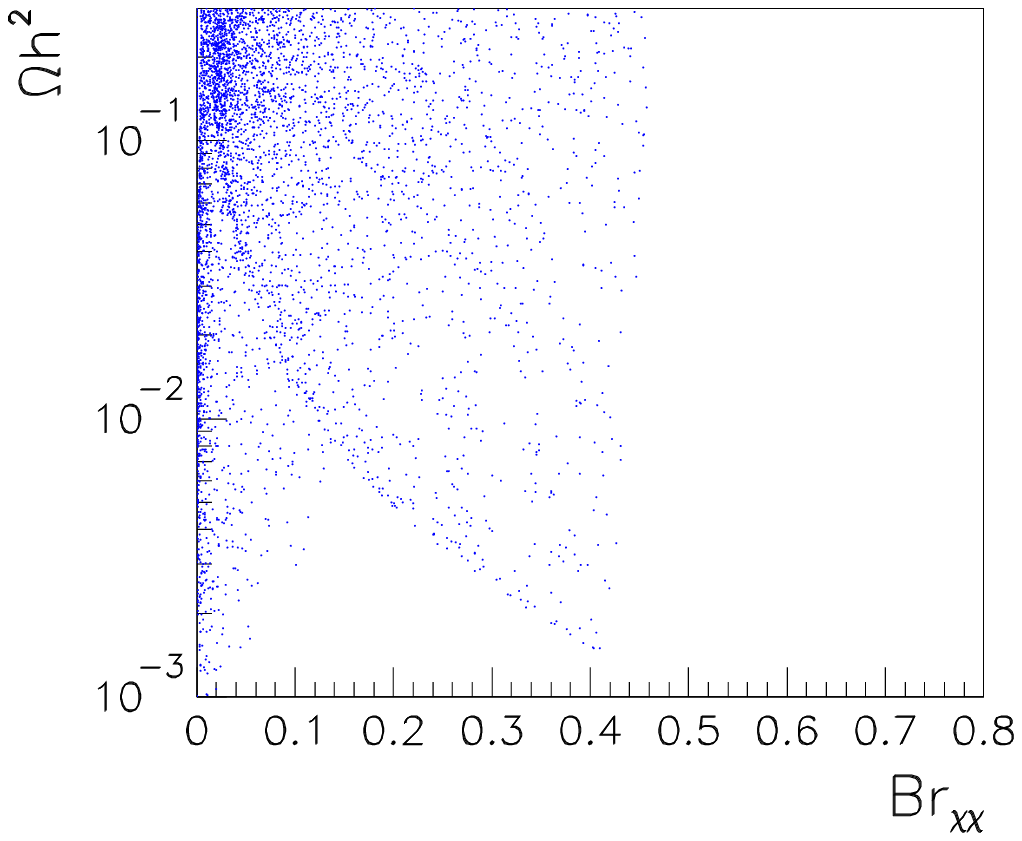}
\caption{\label{tgb10}{\em $\Omega h^2$ {\it vs} $B_{\chi \chi}$
for $\tgb=10$ and $M_1=M_2/5$, scanning over $M_2,\mu$ and $m_0$.
\/.}}
\end{center}
\end{figure*}
For completeness we have also redone the same analysis but with
$\tgb=10$. As expected the largest   $B_{\chi \chi}$  is more
modest than for a lower $\tgb$ and is found to be $45\%$ at most,
as shown in Fig.~\ref{tgb10}. This corresponds to $R_{\gamma
\gamma}>.5$, which means that with enough luminosity, $300fb^{-1}$
at the LHC, one should see the $2\gamma$ signal.

\begin{figure*}[tp]
\begin{center}
\includegraphics[width=16cm,height=10cm]{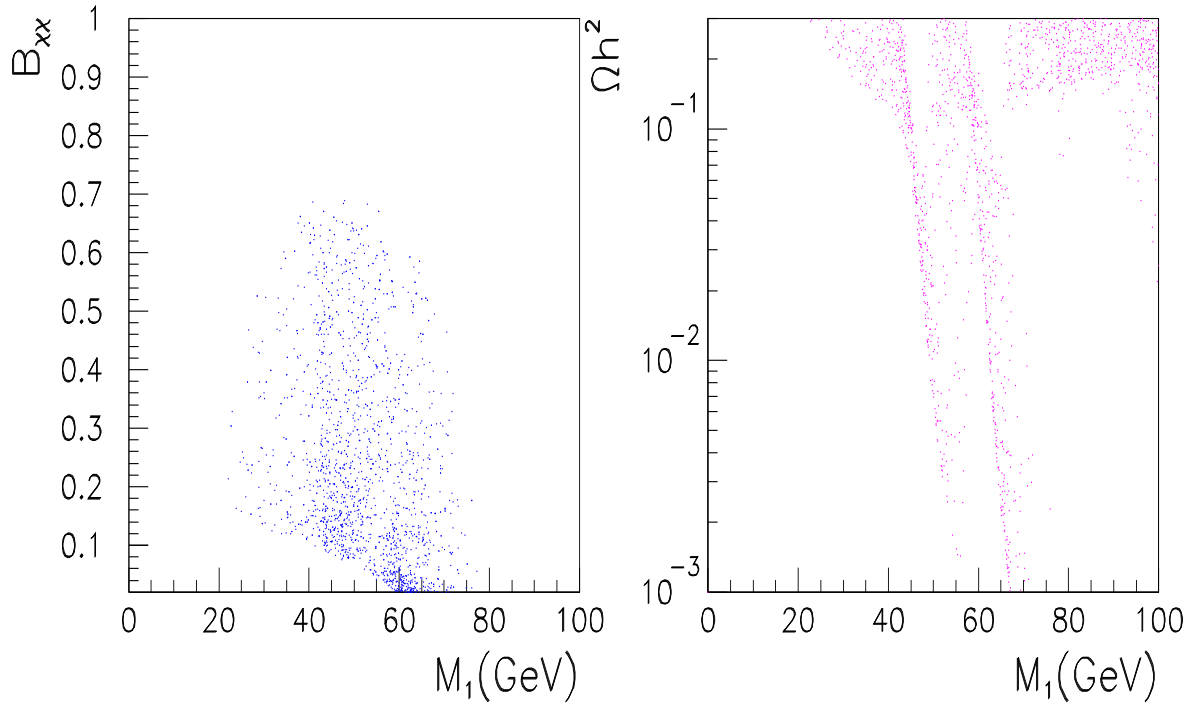}
\caption{\label{tgb5largescan}{\em Large scan over
$M_1,M_2,\mu,m_0$ for $\tgb=5$. The first panel shows the
branching ratio into invisibles {\it vs} $M_1$. The second panel
shows the relic density as a function of $M_1$. Note that one hits
both the $Z$ pole and the Higgs pole. However for the latter
configurations $B_{\chi \chi}$ is negligible.\/}}
\end{center}
\end{figure*}

We have also searched, by making a large scan over $M_1,M_2,\mu$
and  $m_0$, but for fixed $\tgb=5$, which minimum value of $M_1$
one can entertain given our assumption for the slepton spectrum.
Here $M_1$ was varied in the range $10<M_1<100$GeV. We find that,
in order not to have too large a relic density, one can not have
values of $M_1$ below $20$GeV independently of $M_2$ and $\mu$, as
seen in  Fig.~\ref{tgb5largescan}. This is not a value that gives
the largest branching into invisibles since considering the limit
on $\mu$, the mixing is not as strong as with a  value of $M_1$
around $40-50$GeV. Higher values of $M_1$ ($M_1>65$GeV) are safe
since the LSP mass turns out to be too large for the Higgs to
decay  invisibly. Note that values of $M_1 \sim 40-50$GeV within
the gaugino masses unification assumption correspond to $M_2\sim
100$GeV. As we can see from the first panel of Fig.~1, such ``low"
$M_2$ values can only be compatible with the LEP2 limit on the
chargino mass for a large $\mu$ around $400-500$GeV. For such high
$\mu$ there is not enough higgsino component in the LSP. Lower
$\mu$ values require $M_2>180$GeV, ($M_1>90$GeV) leaving no phase
space for the invisible decay of the Higgs. We thus see how useful
it is once again to disconnect $M_1$ from $M_2$.

To conclude we have found that there are still regions of
parameter space that give a substantial branching fraction of the
lightest SUSY Higgs into invisibles that can account both for the
discrepancy in the $g-2$ value and  for the dark matter in the
universe. We also find that these scenarios do not always require
a very light slepton since we can obtain an acceptable amount of
LSP relic density through an efficient annihilation at the $Z$
pole. However scenarios with the largest branching ratio into LSP
do entail that the lightest chargino and at the least the next LSP
are light enough that they could be produced at the Tevatron. The
phenomenology at the Tevatron should somehow be similar to the
Sugra $SU(5)$ based ``24-model" mentioned above and which was
studied in \cite{nonuni-24}. Among other things, due to the fact
that  one has a larger splitting between the LSP and the NLSP, as
compared to the usual unified scenario, one expects an excess of
events containing many isolated leptons originating, for example,
from  a real $Z$ coming from the decay of the NLSP. However to
make definite statements about observability of these states at
the Tevatron requires a thorough simulation. Recently it has also
been pointed out\cite{g-2andLEP} that models with light sleptons
and charginos of a mixed nature (as are required in our analysis
to obtain a large branching into invisibles) apart from helping
give a ``good" $g-2$ at not so large $\tgb$ can also help improve
the $\chi^2$ fits of the electroweak data. It is therefore
important to study in detail phenomenological models out of the
mSUGRA paradigm.

\vspace*{.5cm}

\noi {\bf Acknowledgments}

We would like to thank Sasha~Pukhov  for his collaboration in the
code for the relic density calculation based on {\tt CompHEP}
matrix elements. We also acknowledge a helpful discussion with
Fiorenza Donato regarding the latest BOOMERANG results. We have
also benefited from comments and discussions with the participants
of the Higgs and BSM Working Groups at the {\em Les Houches 2001}
Workshop on {\em Physics at TeV Colliders}. R.G. acknowledges the
hospitality of LAPTH where this work was initiated. This work is
done under partial financial support of the Indo-French
Collaboration IFCPAR-1701-1 {\em Collider Physics}.

\end{document}